\documentclass{IEEEtran}
\usepackage{cite}
\usepackage{amsmath,amssymb,amsfonts}
\usepackage{graphicx}
\usepackage{textcomp,nicefrac}
\usepackage[switch]{lineno}
\def\BibTeX{{\rm B\kern-.05em{\sc i\kern-.025em b}\kern-.08em
T\kern-.1667em\lower.7ex\hbox{E}\kern-.125emX}}
\begin{document}
\title{New software based readout driver for the ATLAS experiment} 
\author{Serguei Kolos, on behalf of the ATLAS TDAQ Collaboration
\thanks{S. Kolos, is with University of California, Irvine, CA 92697-4575, USA (e-mail: serguei.kolos@uci.edu)}
\thanks{Copyright 2020 CERN for the benefit of the ATLAS Collaboration. CC-BY-4.0 license}
}

\maketitle

\begin{abstract}
In order to maintain sensitivity to new physics in the coming years of LHC operations, the ATLAS experiment has been working on upgrading a portion of the front-end electronics and replacing some parts of the detector with new devices that can operate under the much harsher background conditions of the future LHC. The legacy front-end of the ATLAS detector sent data to the DAQ system via so called Read Out Drivers (ROD) - custom made VMEbus boards devoted to data processing, configuration and control. The data were then received by the Read Out System (ROS), which was responsible for buffering them during High-Level Trigger (HLT) processing. From Run 3 onward, all new trigger and detector systems will be read out using new components, replacing the combination of the ROD and the ROS. This  new path will feature an application called the Software Read Out Driver (SW ROD), which will run on a commodity server receiving front-end data via the Front-End Link eXchange (FELIX) system. The SW ROD will perform event fragment building and buffering as well as serving the data on request to the HLT. The SW ROD application has been designed as a highly customizable high-performance framework providing support for detector specific event building and data processing algorithms. The implementation that will be used for the Run 3 is capable of building event fragments at a rate of 100 kHz from an input stream consisting of up to 120 MHz of individual data packets. This document will cover the design and the implementation of the SW ROD application and will present the results of performance measurements performed on the server models selected to host SW ROD applications in Run 3.
\end{abstract}


\section{Introduction}
\label{sec:introduction}
As part of the preparation for LHC Run 3, which will begin at the end of 2021, the ATLAS experiment \cite{Ref-ATLAS} has upgraded some parts of the detector with new components that can operate under the much harsher background conditions expected as the LHC reaches higher instantaneous luminosity. The new detector and trigger systems will use modern Front-End (FE) electronics that require an updated readout system.  The legacy FE of the ATLAS detector sent data to the TDAQ  system \cite{Ref-TDAQ} via so-called Read Out Drivers (ROD) \cite{Ref-ROD} - custom made VMEbus boards devoted to data processing, configuration and control. These data were then sent to the Read Out System (ROS) \cite{Ref-TDAQ}, that was responsible for buffering and serving them to the High-Level Trigger (HLT) \cite{Ref-TDAQ}. From Run 3 onward, all new trigger and detector systems will be read out using new components, replacing the combination of the ROD and the ROS. This  new path will feature a new facility called the Software Read Out Driver (SW ROD), which will receive FE data via the FE Link eXchange (FELIX) system \cite{Ref-FELIX}, perform event fragment building and buffering as well as serving data on request to the HLT. 

\section{FELIX System Overview}
FELIX is a new generic detector readout system that can receive data from detector FE electronics via (among others) the versatile radiation hard optical link architecture \cite{Ref-Versatile} developed at CERN. FELIX can be used to receive data either via GigaBit Transceiver (GBT)\cite{Ref-GBT} or the in-house designed FULL mode protocol. FELIX uses a custom PCIe card that receives data via optical links and passes them to the memory of a commodity computer via PCIe bus. FELIX also provides a software application that can forward these data to a number of subscribers via a commercial network using Remote Direct Memory Access (RDMA) to maximize performance. RDMA is a technology that makes it possible to put data directly into the main memory of another computer without involving the processor, cache or operating system of that computer. FELIX implements a custom network communication layer called NetIO \cite{Ref-NetIO} on top of the RDMA over Converged Ethernet (RoCE) protocol that is supported by many modern network cards. NetIO provides a C API that can be used by a software application to receive data from the FELIX system. A FELIX card can be operated in two modes using the respective protocols:
\begin{enumerate}
\item GBT Mode: with the GBT protocol a physical input link can be subdivided into a number of logical sublinks (known as E-Links), which can pass information from separate pieces of FE electronics. For Run 3 the maximum number of E-Links for a single FELIX card is limited to 192, which in this case are equally spread over 24 GBT links.
\item FULL Mode: this mode has no logical subdivision of links and uses an in-house designed protocol for higher bandwidth. For Run 3 this mode can be used to send data either via 12 links at full occupancy with the speed of 9.6 Gbps or via 24 links with 50\% occupancy (4.8 Gbps).
\end{enumerate}

\section{SW ROD System Architecture}
\begin{figure}[!htbp]
\centerline{\includegraphics[width=3.5in]{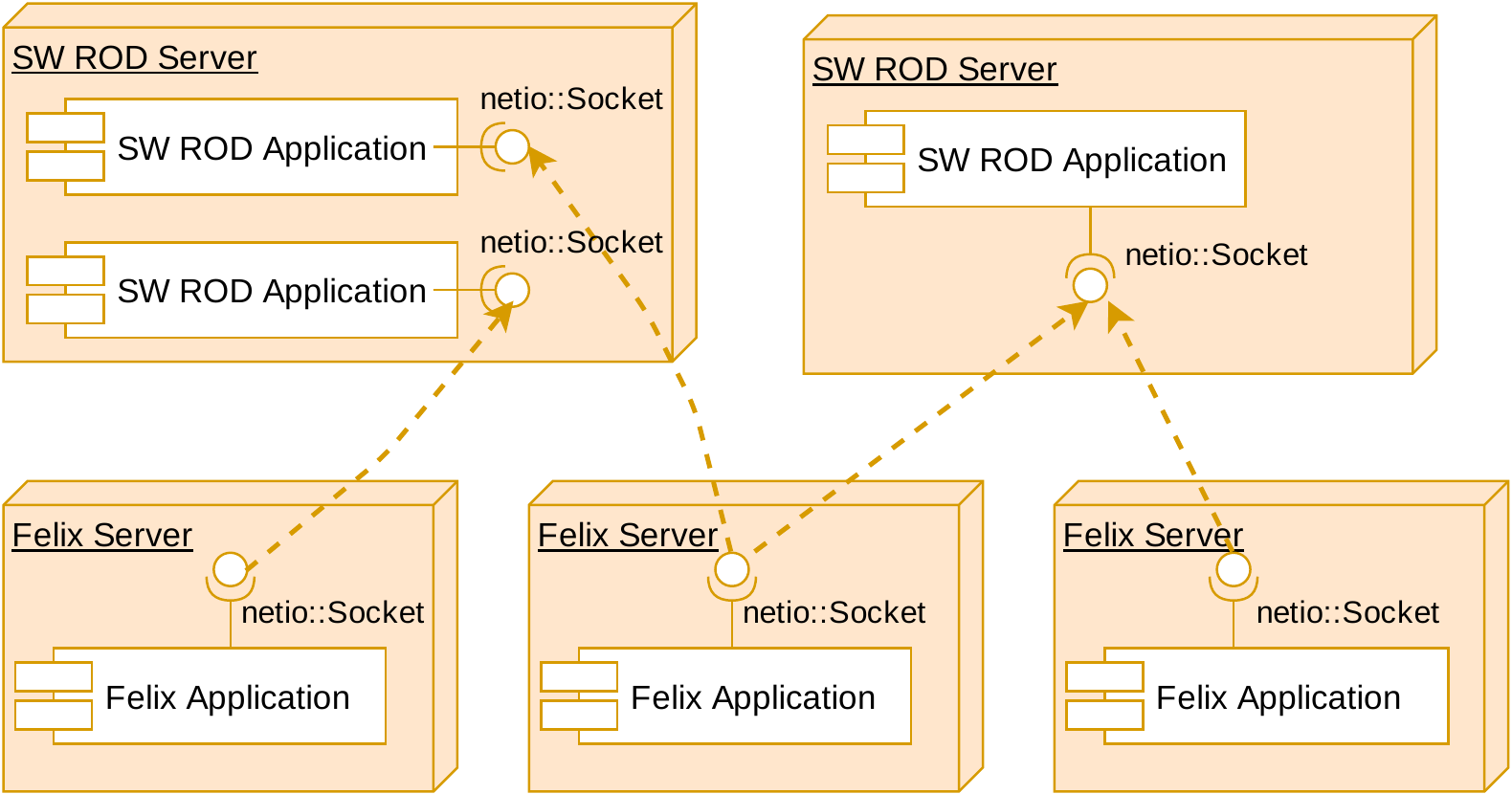}}
\caption{SW ROD deployment.}
\label{fig1}
\end{figure}

The SW ROD facility is envisaged to be implemented as software running on a set of commodity computers. Given that a single computer can serve only a limited amount of input data, and in order to scale to the size of the new ATLAS readout system, the software had to be designed in a way that allowed it to be distributed over an arbitrary number of computers. In the current design this is achieved by splitting the input data channels between a number of software processes, which are referred to as SW ROD applications as shown in \figurename~\ref{fig1}. 

Each instance of the SW ROD application can run on a separate computer, but it originates from the same binary executable. This executable implements a highly customizable high-performance framework, rendering support for detector specific event building and data processing algorithms provided in the form of shared libraries (a.k.a. binary plugins). This way different instances of the SW ROD application diverge by using distinct configurations that define a set of plugins to be used as well as their configuration parameters. 

\section{SW ROD Application Functional Requirements}
The Read Out Drivers being used by the legacy readout system to receive and process data from the ATLAS detector FE were developed independently by every subdetector. As such, they perform subdetector specific data processing and event building based on the signals received from the ATLAS Central Trigger Processor (CTP) \cite{Ref-CTP}. As the FELIX system does not perform any data processing or event aggregation, but merely provides data routing between detector FE and the DAQ system, the task of data aggregation and processing has to be fulfilled by the SW ROD application before transferring data to the HLT farm.  It is also expected that the SW ROD application will be used not only for normal physics data taking but for various auxiliary subdetector specific activities, such as commissioning, calibration, monitoring, debugging etc., in which data would have to undergo specific processing and may need to be transferred to a different destination than the HLT farm. 
To meet such requirements the SW ROD application has been designed as a framework that supports a high degree of customization by making it possible to load subdetector specific event building and data processing algorithms at run time, which can be further configured by subdetectors with respect to their specific needs.

\section{SW ROD Application High-Level Design}

\begin{figure}[!htbp]
\centerline{\includegraphics[width=3.5in]{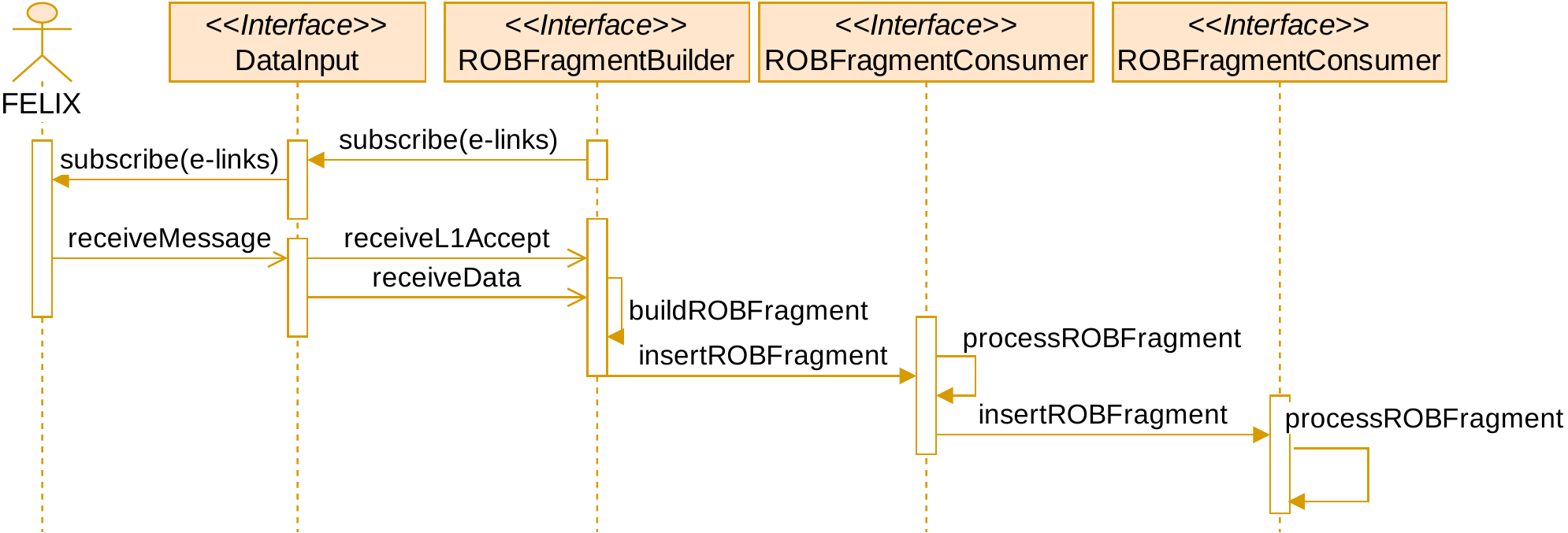}}
\caption{Typical interactions between SW ROD application components for a normal data taking activity. The DataInput component subscribes to FELIX and passes received data to the ROBFragmentBuilder, which aggregates data into event fragments and transfers them to the first ROBFragmentConsumer in the list. }
\label{fig2}
\end{figure}

The SW ROD application is split internally into a number of independent components, with each of them providing a simple interface that defines how other components can interact with it. There are three main components defined by the SW ROD application architecture that can be interacted via the respective interfaces:

\begin{itemize}
\item DataInput interface: abstracts a source of input data to shield the other components of the SW ROD application from any changes in the network input protocol. In addition it also makes it possible to use another data source, for example internal data generators, for testing and debugging.

\item ROBFragmentBuilder interface: abstracts implementations of data aggregation algorithms, which may need to facilitate different data handling strategies as required by subdetectors and should be able to support different FELIX operation modes. This approach scales well with a number of data aggregation algorithms, as adding a new algorithm does not require modification of the existing ones. It also offers the possibility to support auxiliary data aggregation strategies to be used for calibration or monitoring without affecting the basic procedures used for normal physics data taking. An implementation of this interface is responsible for aggregating data chunks from individual E-Links received via the DataInput interface into event fragments according to the given configuration. Such a configuration defines the set of event fragments to be produced as well as a list of input links for each fragment.

\item ROBFragmentConsumer interface: abstracts any kind of processing that can be applied to fully aggregated event fragments. Multiple implementations of this interface can be used simultaneously in the same SW ROD application, in which case they will be organized into a singly-linked list. Each consumer in this list will have to forward event fragments to the next one after finishing its specific processing step. For example, as shown in \figurename~\ref{fig2}, one implementation of this interface can apply a custom subdetector specific processing procedure to the event fragments before passing them to another consumer that is used to transfer these fragments to the HLT farm.
\end{itemize}

As shown in this diagram the data handling is done by the implementations of the SW ROD application interfaces, while the Application itself merely loads and instantiates the corresponding implementation classes in accordance with a given configuration and links the instantiated objects in the order defined by this configuration. 

\section{SW ROD Default Component Implementations}
\begin{figure}[!htbp]
\centerline{\includegraphics[width=3.5in]{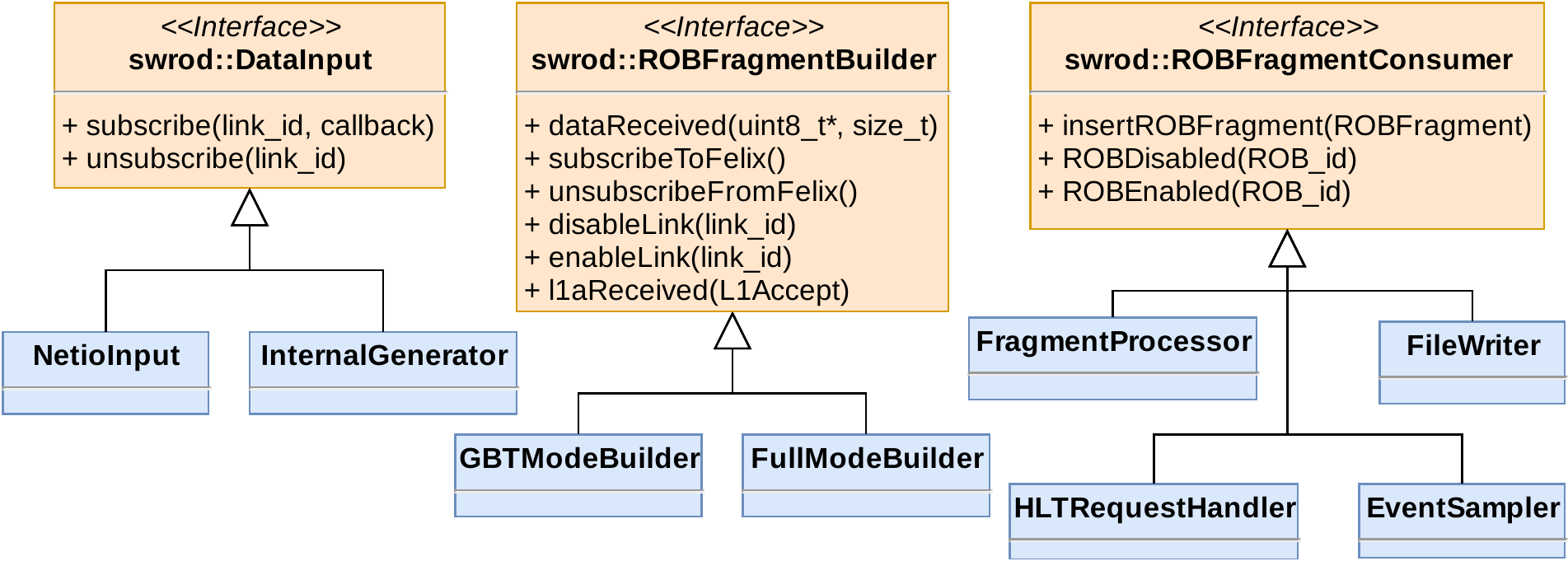}}
\caption{Default SW ROD interface implementations.}
\label{fig3}
\end{figure}

A shared library that contains default implementations for all three interfaces is supplied along with the SW ROD application. This library contains all classes shown in \figurename~\ref{fig3}.

\subsection{DataInput Interface Implementations}

\begin{itemize}
\item The NetioInput class is responsible for receiving data from the FELIX system using the NetIO Socket interface for a given set of E-Links and passing these data to the fragment builder via the ROBFragmentBuilder interface. 
\item The InternalDataGenerator can generate FELIX-like data chunks of a given size for a configurable number of E-Links. This class is used for debugging as well as for unit test implementation.
\end{itemize}

\subsection{ROBFragmentBuilder Interface Implementations}
The library provides two implementations of the ROBFragmentBuilder interface, which can be used to receive data from the FELIX system in either GBT or FULL mode. The algorithms implement a specific data aggregation strategy in a generic way that is independent of the format of the incoming data chunks. As this format is detector specific this feature was implemented by allowing detectors to supply two custom procedures as parameters for these algorithms: 

\begin{itemize}
\item Trigger Information Extraction procedure - this is a function that extracts the Level 1 Trigger identifiers from a given data chunk. These identifiers are used to assign data chunks to a particular event fragment as well as to align data with the Trigger information received from the CTP. 

\item Data Integrity Checking procedure - this function is intended to be used if there is a suspicion that input data chunks could be corrupted or a sequence of data packets for a particular input link is broken. This function is assumed to know the location of the checksum value in a given data packet format as well as the Cyclic Redundancy Check (CRC) algorithm that was used to calculate that value.
\end{itemize}

In most cases detector developers have only to define these functions and reuse the data aggregation strategies provided by the carefully optimized and extensively tested default ROBFragmentBuilder interface implementations. On the other hand, if another event fragment aggregation strategy is required for a particular subdetector, a new algorithm can be implemented and plugged in to the SW ROD application as is done for the default implementations. This does not affect the existing components of the SW ROD application and is completely transparent for the application itself.

\subsection{ROBFragmentConsumer Interface Implementations}

\begin{itemize}
\item The FragmentProcessor class was developed to simplify implementation of the common task, required by many subdetectors, of applying custom detector-specific post-processing to all event fragments produced by the given SW ROD application. This class provides a workbench to execute detector-specific code on every event fragment that is passed to this consumer. This code should perform the necessary modifications to the event fragment payload but should keep the structure of the fragment untouched. The code can be provided in the form of a function, which has to be implemented by the corresponding detector experts and be given to the SW ROD application in the form of a shared library that will be loaded at runtime. 

\item The HLTRequestHandler class is responsible for buffering event fragments and serving them to the HLT farm on request.  It keeps the event fragments until informed by the HLT that they are no longer needed. Event fragments are indexed by their Level 1 Trigger identifier and stored in an internal buffer until a clear request has been received from the HLT. On receipt of a clear request, all the event fragments with the identifiers provided by this request will be removed from the index and their allocated memory freed.

\item The FileWriter class implements a consumer that simply writes all received event fragments to a file on disk. Files created by the FileWriter will be in the standard ATLAS data file format \cite{Ref-Eformat} with all event fragments prepended by the ATLAS full event header, which make such files compatible with standard ATLAS event processing and analysis applications. This functionality is useful for testing, commissioning, calibration and other auxiliary activities which are performed by detectors beyond normal data taking. 

\item The EventSampler implements event selection for online monitoring. An instance of this class can be optionally added to the list of a SW ROD application consumers to select a subset of aggregated event fragments for the purpose of online monitoring. This class passes selected events to the TDAQ Event Monitoring service \cite{Ref-Emon} that transfers them to the applications responsible for data quality assessment.
\end{itemize}

\section{SW ROD Application Performance Requirements}
In Run 3, the SW ROD has to be able to operate at an input rate of 100 kHz, matching the ATLAS Level 1 Trigger accept rate. The number of input links and the overall data rates are defined by the output produced by the FELIX system. Table~\ref{tab1} summarizes these numbers for a single FELIX card.

\begin{table}
\centering
\caption{FELIX Card output rates for Run 3}
\setlength{\tabcolsep}{3pt}
\begin{tabular}{|p{20pt}|p{25pt}|p{40pt}|p{42pt}|p{40pt}|p{35pt}|}
\hline
Mode & Chunk Size (B) & 
Chunk Rate per Link (kHz) & 
Links per FELIX Card & 
Chunk Rate per Card (MHz) & 
Data Rate per Card (GB/s)\\
\hline
GBT& 40& 100& 192& 19.2& 0.77 \\
\hline
FULL& 5000& 100& 12 (24)& 1.2 (2.4)& 6 \\
\hline
\end{tabular}
\label{tab1}
\end{table}

An important goal of the SW ROD is to handle as many input data links as possible in order to reduce the total system cost by minimizing the number of computers to be used to run SW ROD applications. While in FULL mode this number is essentially limited by the input network bandwidth available for a SW ROD computer, in GBT mode the data rate produced by a single FELIX card is much lower and the number of input links that can be served by a single SW ROD computer is instead limited by the performance of the GBT data aggregation algorithm. A dedicated study has been performed to estimate the maximum number of input links that can be handled by the GBT event fragment aggregation algorithm executed on a single SW ROD computer. The results of this study will be presented in the next section.

\section{GBT Mode Event Fragment Building Algorithm Optimization}
Due to power consumption and heat dissipation issues the clock frequency of a modern CPU is normally in inverse proportion to the number of cores for the given CPU. Thus the product of these parameters gives a similar value for any CPU in the same price range. This value can be used to make a rough estimate of the full computing power a particular CPU can offer. It should be noted that a modern CPU is capable of executing more than one operation per cycle, but in practice this is difficult to achieve for complex code and normally a one-to-one ratio between cycles and operations is considered satisfactory. Taking 2.5 GHz as an average CPU frequency for a CPU that has 10 cores, one can assess the total number of CPU operations per second provided by an averagely priced CPU to be on the order of $2.5\cdot10^{10}$.

Given that the rate of data chunks from a single FELIX card in GBT mode is about 20 MHz, a simple division shows that such a CPU can provide about 1200 operations for a single data chunk, which corresponds to 0.5 microseconds. It should be taken into account that every chunk has to be aligned, by means of the Level 1 Trigger identifier that every data chunk contains, with the other ones for the purpose of event fragment aggregation. From this point of view this amount of computing power does not look large, especially if one wants to maximize the number of FELIX cards that can be handled by the same SW ROD computer, in which case this budget has to be divided further accordingly. Moreover, the computational resources provided by a modern CPU are proportional to the number of CPU cores, which means that to utilize them in an efficient way the software has to be designed to use multiple threads with a high degree of parallelism. This essentially precludes the use of high-level design patterns, like producer-consumer, to pass data between threads as this would incur too much performance overhead for thread synchronization. 

The solution that was implemented for the GBT event fragment aggregation algorithm to minimize the rate of interactions between threads was to combine both data reading and event fragment aggregation into the same threads. To achieve that the total number of input E-Links is split among a configurable number of worker threads, with each thread reading data chunks from the given subset of E-Links and aggregating them into a subfragment of a given event. When all subfragments of a particular event are ready they are assembled together by a dedicated fragment building thread. This approach makes it possible to split the algorithm into two stages: 

\begin{enumerate}
\item The processing of individual data chunks is done in parallel by multiple concurrent threads at the O(10) MHz rate.
\item The final event fragment assembly that requires synchronization between threads is done at the rate of 100 kHz only.
\end{enumerate}

The degree of parallelism provided by this algorithm can be estimated using a formula that is based on Amdahl's law: 

\begin{equation}
\label{eq1}
S(n)=\frac{1}{(1-P) + \frac{P}{n}}.
\end{equation}
 
Equation (\ref{eq1}) defines how the speedup $S(n)$ of an algorithm executed by a given number of threads $n$ depends on the parallel fraction of this algorithm $P$. Given that we know the processing rates of the parallel and non-parallel fractions of the GBT event fragment aggregation algorithm we can express $P$ as (\ref{eq2}).

\begin{equation}
\label{eq2}
P=1 - C\textsc{fa} \times \frac{10^5}{10^7} = 1 - 0.01 \times C\textsc{fa}.
\end{equation}

Here $C\textsc{fa}$ is the relative cost of the final subfragment assembly operation with respect to the cost to handle a single data chunk. This equation shows that if the relative cost of the final assembly operation is less than 100 then the algorithm should give some performance gain, but in order to scale well this number should be at least less than 50. Using this equation (\ref{eq1}) can be transformed to: 

\begin{equation}
\label{eq3}
S(n)=\frac{n}{0.01 \times C\textsc{fa} (n-1) + 1}.
\end{equation}

This equation defines how the speedup of the GBT algorithm depends on the relative cost of the final assembly operation. Finally, inverting (\ref{eq3}) yields (\ref{eq4}), which will be used in the next chapter to assess $C\textsc{fa}$ for the current algorithm implementation using the empirical values for $S(n)$ obtained from performance measurements:

\begin{equation}
\label{eq4}
C\textsc{fa}=\frac{(\frac{S(n)}{n} - 1)}{0.01 \times (n-1)}.
\end{equation}

\section{Performance Measurements}
\subsection{Testbed Configuration}
Event building algorithm performance measurements were performed on a testbed that replicates the same hardware configuration that will be used by the readout system during Run 3:
\begin{itemize}
\item SW ROD application running on a computer with a dual-socket Supermicro motherboard with 2 Intel(R) Xeon(R) Gold 5218 CPUs and 96 GB of DDR4-2667 RAM. Each CPU has 16 physical cores with a base frequency of 2.3~GHz. 
\item Input data for the tests generated by a FELIX card software emulation application running on another computer with an Intel Xeon E5-1660 v4 CPU with 3.2 GHz base frequency and equipped with 32 GB DDR4 2667 MHz memory. 
\item Both computers were equipped with Mellanox ConnectX-5 100 GbE network adapters, which were connected via a 100 Gb network switch. Data were sent to the SW ROD application via the FELIX NetIO protocol. 
\end{itemize}

\subsection{Network Throughput Test}
In order to access the overhead of the RoCE protocol the network throughput was measured using a simple bandwidth test utility from the Mellanox OFED software package with a default packet size of 65K bytes. The receiving application was started on the SW ROD computer with the following command:

\verb|# ib_send_bw -F  -n 100000|

The client (sending) application was started on the FELIX computer with the IP address of the SW ROD host:

\verb|# ib_send_bw -F  -n 100000 192.168.100.1|

Both applications reported an average rate of 91.3 Gb/s that stayed almost constant throughout the test, with marginal variations of less than a fraction of 1 Gb/s.

\subsection{GBT Mode Tests}
\begin{figure}[!htbp]
\centerline{\includegraphics[width=3.5in]{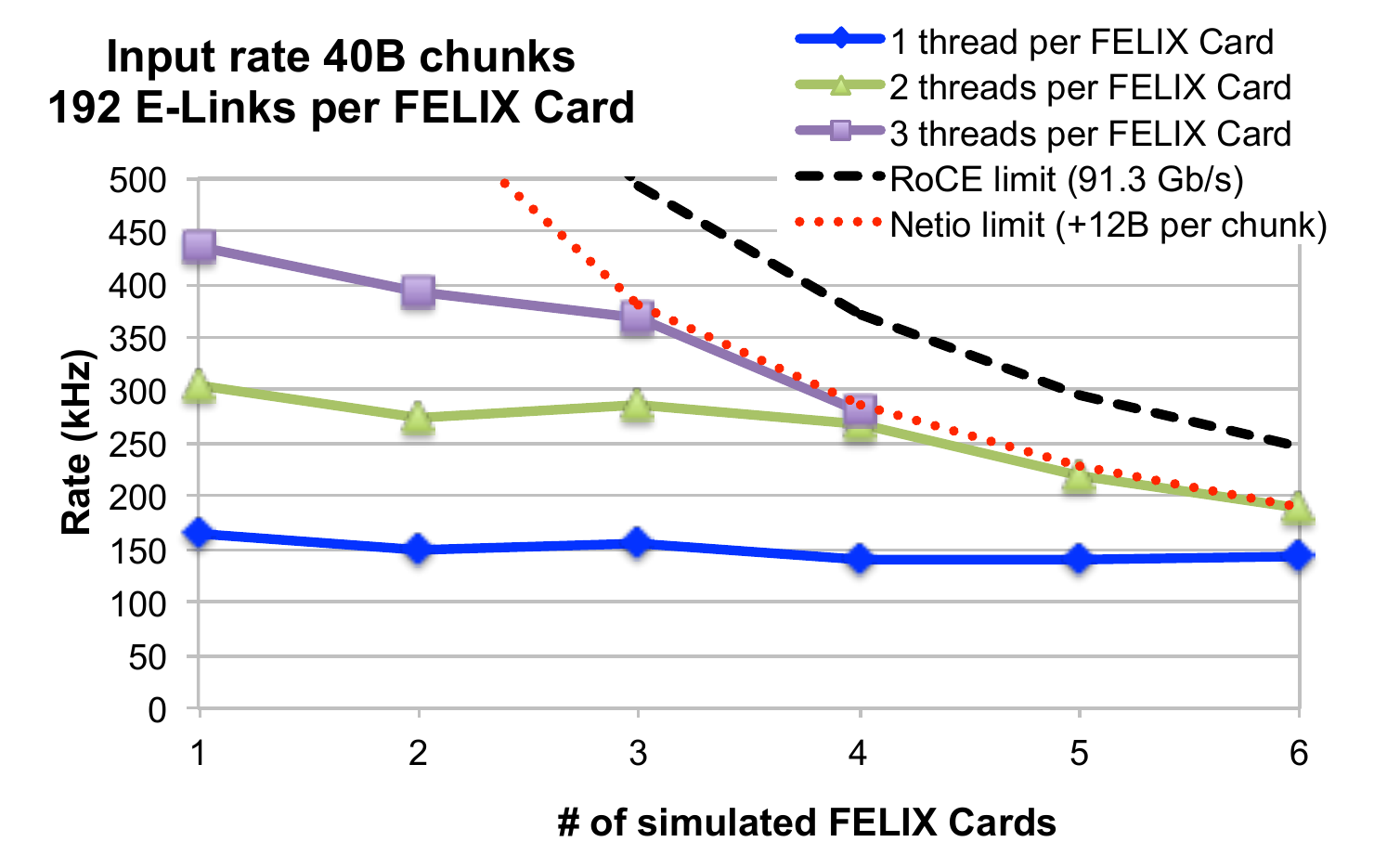}}
\caption{GBT event fragment aggregation algorithm performance.}
\label{fig4}
\end{figure}

The aim of these tests was to study how the GBT event fragment building algorithm scales with the number of input E-Links and the number of threads used to handle input data. To this end, three series of tests were performed with the SW ROD application using one, two and three threads respectively to receive and aggregate data chunks from every group of 192 input links, which corresponds to input from a single FELIX card. The total number of emulated FELIX cards for different test series varied from 1 to 6, which made for a total number of input channels increasing gradually from 192 to 1152.  The size of the generated data chunks was set to 40 bytes. The results of these tests are shown in \figurename~\ref{fig4}.

These results show that the GBT event fragment aggregation algorithm implementation scales well in both dimensions: with the number of worker threads aggregating data from a given number of E-Links as well as with the number of such aggregation operations running concurrently in the scope of the same SW ROD application. 

The dotted line shows the maximum theoretical input rate that can be obtained with the given hardware configuration, which is limited by the available network bandwidth. This line represents (\ref{eq5}):

\begin{equation}
\label{eq5}
L=\frac{91.3 \times 10^9}{192 \times F \times (40*8+12*8)},
\end{equation}

where $L$ is the input rate, $F$ is the number of simulated FELIX cards, $40 \times 8$ is the size of the data chunk in bits, $12 \times 8$ is the size of the NetIO protocol overhead per chunk in bits as well and $91.3 \cdot 10^{9}$ is the maximum bandwidth that can be achieved with using the RoCE protocol in Gb/s. This line demonstrates that the last three results of the tests with two reading threads and all but the first two results for the test series with three reading threads were limited by the network bandwidth. The dashed line shows the maximum rate that could be achieved if the NetIO protocol overhead was equal to zero. It indicates that the input rate could potentially be improved by reducing the NetIO overhead.

Using the results which were not limited by network bandwidth one can calculate the speedup $S(n)$ and parallel fraction $P$ of the GBT event fragment assembly algorithm and then use these numbers with (\ref{eq4}) to compute an estimate of the $C\textsc{fa}$ coefficient. The results of these calculations are shown in Table~\ref{tab2}.

\begin{table}
\centering
\caption{Estimate of the parallel fraction of GBT algorithm}
\setlength{\tabcolsep}{3pt}
\begin{tabular}{|p{40pt}|p{50pt}|p{30pt}|p{30pt}|}
\hline
N of threads &  Average $S(n)$ & $C\textsc{fa}$ & $P$ \\
\hline
2 & 1.86 &  7.5 & 0.925\\
\hline
3 & 2.65 &  6.6 & 0.934\\
\hline
\end{tabular}
\label{tab2}
\end{table}

\subsection{FULL Mode Tests}
In FULL mode, larger sized event fragments are sent to FELIX over fewer (higher bandwidth) links, with no data aggregation required in the SW ROD. The number of links needing to be serviced at this increased bandwidth can vary from 1 to 24 in the extreme case. In FULL mode several links can be grouped together to be used to send fragments corresponding to different Level 1 trigger events in a round-robin pattern from the same piece of detector FE electronics if more bandwidth than a single link can provide is required.

\begin{figure}[!htbp]
\centerline{\includegraphics[width=3.5in]{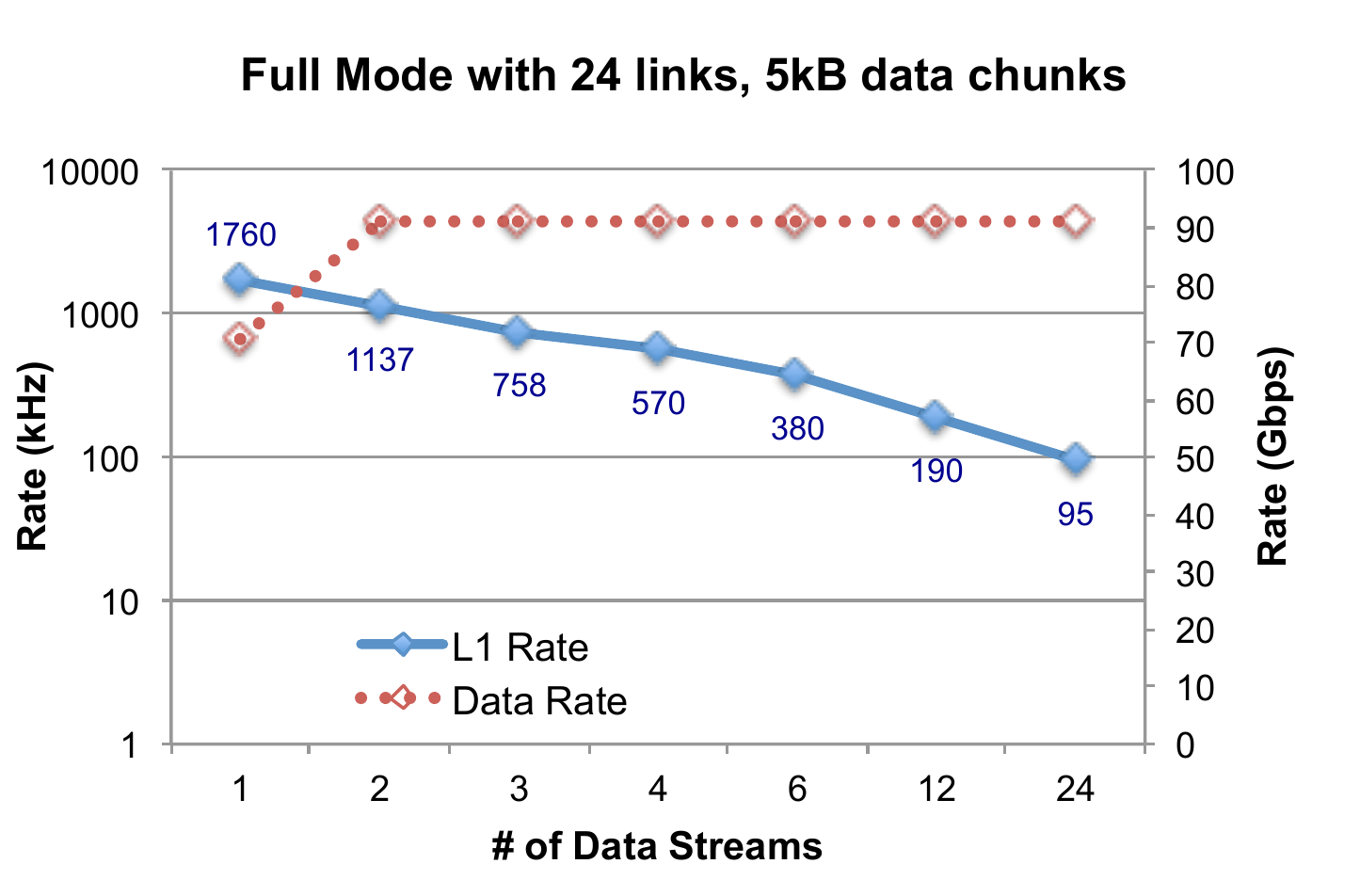}}
\caption{FULL mode data handling algorithm performance.}
\label{fig5}
\end{figure}

Tests have been performed to study the behavior of the SW ROD's FULL mode data handling algorithm relative to the number of input link groups, which need to be serviced independently. Each group of links was used to send an independent stream of data and inside a group event fragments were sent over the given links using round-robin pattern. For these tests the size of the generated packets was set to 5K bytes and the number of independent streams of data generated by the FELIX software simulator varied from 1 to 24. For each configuration the average input rate per data stream was measured. The results of these tests are shown in \figurename~\ref{fig5}.

The results demonstrate the excellent scalability of the FULL mode data handling algorithm with respect to the number of input links. In all test series except one input rate was limited by the network bandwidth. The only exception is the configuration with all 24 input links used for the same data stream. In this test the input rate went up to 1.76 MHz, which saturated the CPU cores used by the SW ROD application's reading threads. No further study has yet been done for this scenario as the rate that was achieved is far in excess of the readout performance requirements for Run 3.

\subsection{Scalability Towards Run 4 Requirements}
For Run 4, which is planned to start in 2027, the LHC will undergo the High Luminosity Upgrade \cite{Ref-HLLHC} that will significantly increase instantaneous luminosity and the number of particle interactions per bunch crossing. This will bring new requirements for the ATLAS TDAQ system, which will have to receive data from trigger and detector electronics at an input rate of 1 MHz. As the readout system for Run 4 will be based on FELIX it is useful to study the limits of the current readout implementation, such that they can be addressed in the new TDAQ architecture. For this reason, a series of tests were performed with the GBT event fragment aggregation algorithm to reveal the maximum number of input E-Links and chunk size configurations at which 1 MHz input can be sustained. 
For these tests the SW ROD application used the default GBT event fragment aggregation algorithm that assembles data from all the given input E-Links to a single fragment. Two rounds of tests were performed. In the first one the number of input E-Links varied from 24 to 192 and the input data chunk sizes were set to 40 and 80 bytes for different test series. For the second round of tests the number of input E-Links was fixed to be 48 and the data chunk size varied from 40 to 240 bytes. The algorithm used 6 reading threads for both rounds of tests.

\begin{figure}[t]
\centerline{\includegraphics[width=3.5in]{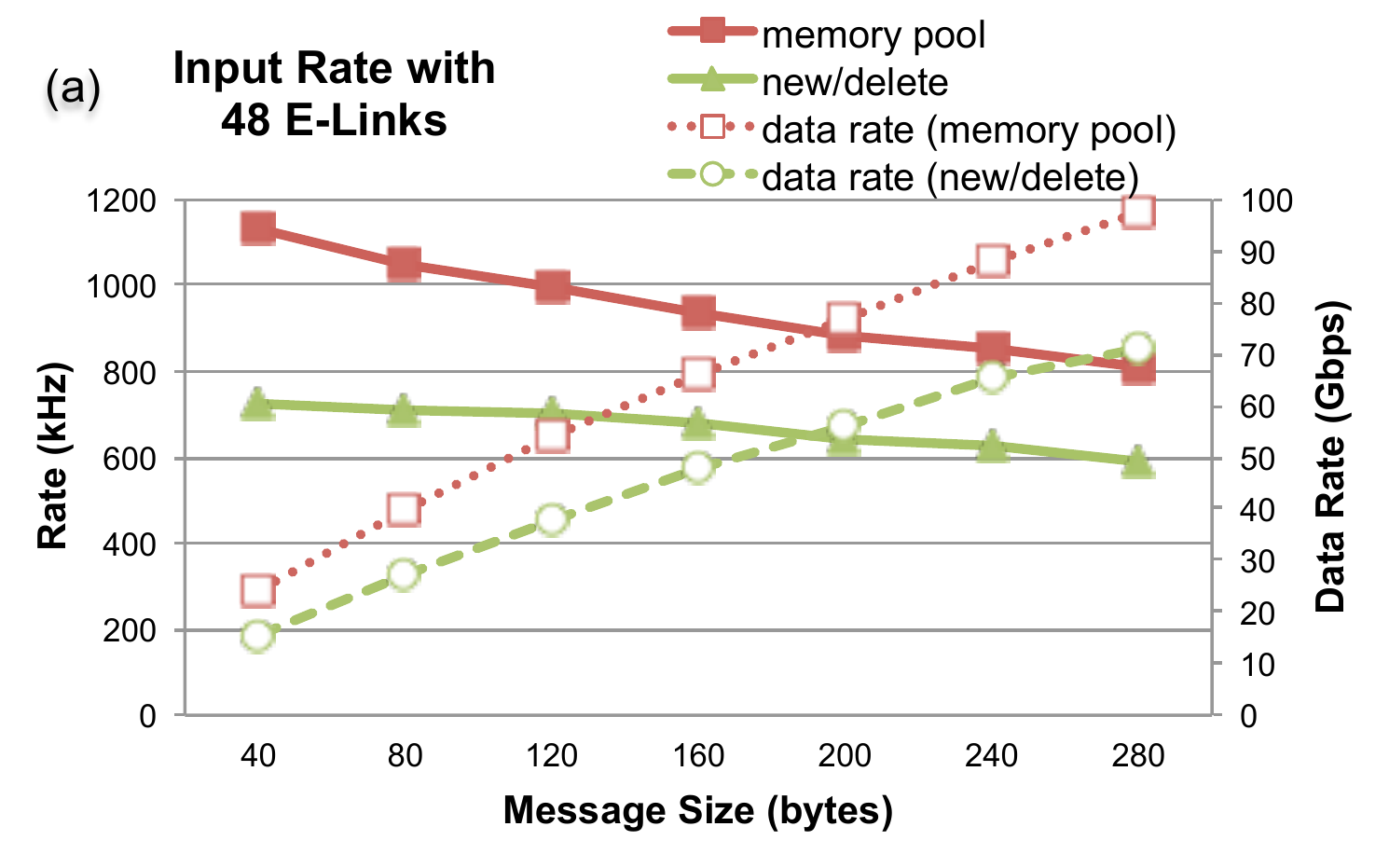}}
\centerline{\includegraphics[width=3.5in]{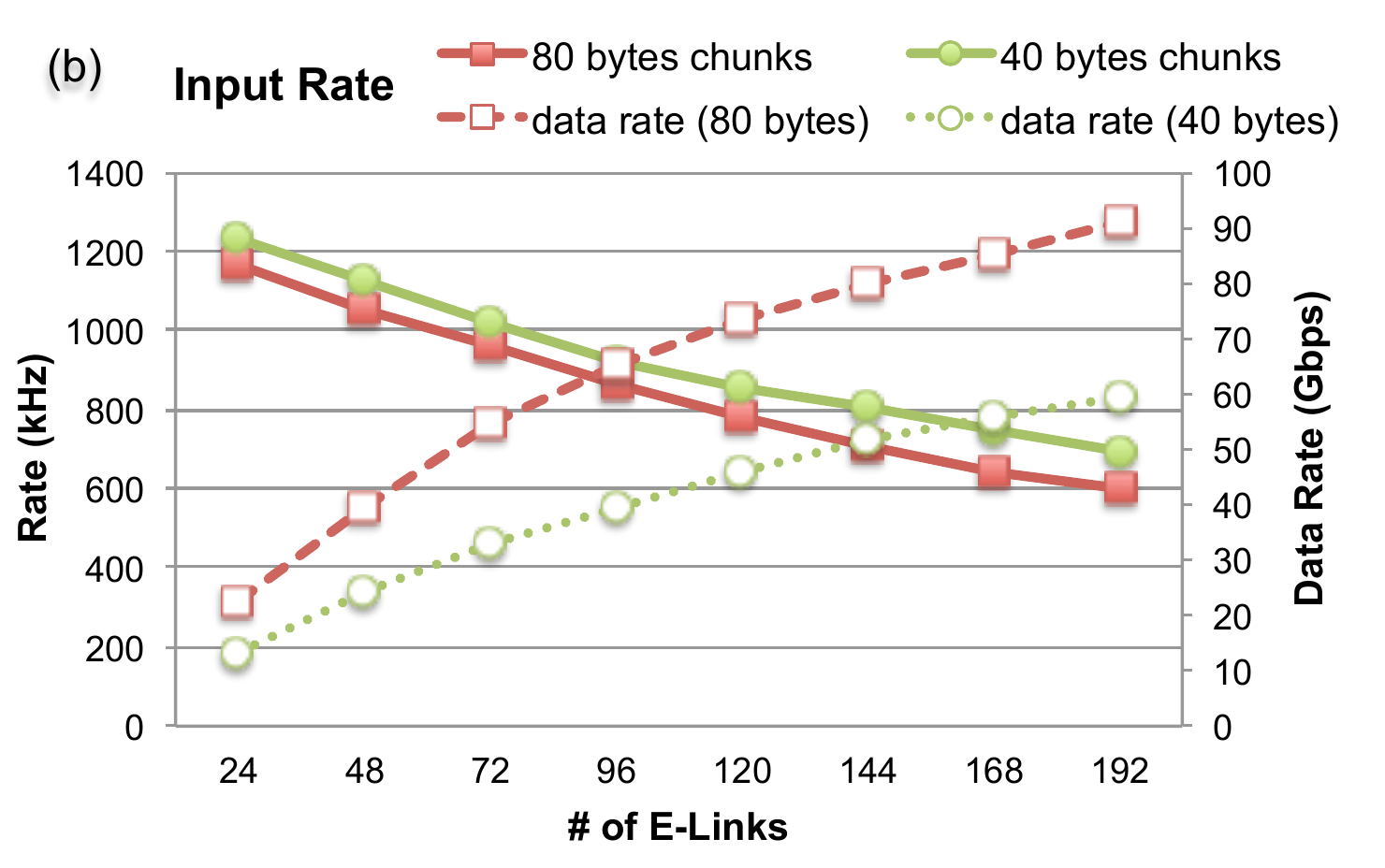}}
\caption{SW ROD input rate with varying data chunk size (a) and with varying number of E-Links (b).}
\label{fig6}
\end{figure}

\figurename~\ref{fig6}a shows two data series which were obtained with the same test configurations but using  different versions of the SW ROD application. The first test series revealed a bottleneck in the SW ROD application that was caused by the standard new and delete memory management operations. This was not a problem for the previous tests, where these operations were taking place at a rate of about 100 kHz, but when the input rate was increased towards 1 MHz the memory management overhead became prominent. A quick solution was put in place by replacing the new and delete operations with a custom memory pool implementation that preallocates a large number of memory blocks and keeps a list of free blocks in a \verb|tbb::concurrent_queue| container \cite{Ref-TBB}, which made the use of this memory pool by multiple concurrent threads possible. This improved the input rate of the SW ROD application by almost 50\% and made it possible to reach a rate above 1 MHz with some configurations. The same implementation was used for the second round of tests, for which the results are shown in \figurename~\ref{fig6}b.

\section{Conclusion}
A mixture of the legacy ROD-based and the new FELIX-based readout will be used by the ATLAS TDAQ system for LHC Run 3. The SW ROD is a new component of the ATLAS DAQ system that was developed to receive data from the FELIX. The SW ROD implements a high performance customizable framework that supports custom input data formats and different event fragment aggregation strategies as required by the new ATLAS detector and trigger components. The SW ROD fully satisfies the performance and functional requirements which have been defined by ATLAS for Run 3. The default GBT event fragment aggregation algorithm makes it possible to handle data input at or above the required rates from up to 6 FELIX cards working in GBT mode, thus minimizing the overall cost of the new readout system by reducing the number of required computers for the SW ROD system. Further optimisation could be achieved by reducing the overhead of the FELIX communication protocol. A study of how the Run 4 performance requirements can be met is ongoing and has already revealed some very promising results.

\bibliographystyle{IEEEtran}
\bibliography{Paper}

\end{document}